\documentclass[aps,twocolumn,prd,nofootinbib,floatfix,showpacs]{revtex4}
\usepackage[dvips]{graphicx}
\usepackage{amsmath}
\usepackage{amssymb}
\usepackage{bm}
\newcommand{\vek}[1]{\bm{\mathrm{#1}}}
\begin{document}


\title{Surface effects in color superconducting strange-quark matter} 


\author{Micaela Oertel}
\affiliation{LUTH, Observatoire de Paris, CNRS, Universit\'e Paris
  Diderot, 5 place Jules Janssen,
  92195 Meudon, France}
\author{Michael Urban}
\affiliation{IPN Orsay, CNRS-IN2P3 and Universit\'e Paris-Sud,
  91406 Orsay cedex, France}


\begin{abstract}
  Surface effects in strange-quark matter play an important role for
  certain observables which have been proposed in order to identify
  strange stars, and color superconductivity can strongly modify these
  effects. We study the surface of color superconducting strange-quark
  matter by solving the Hartree-Fock-Bogoliubov equations for finite
  systems (``strangelets'') within the MIT bag model, supplemented
  with a pairing interaction. Due to the bag-model boundary condition,
  the strange-quark density is suppressed at the surface. This leads
  to a positive surface charge, concentrated in a layer of $\sim$ 1 fm
  below the surface, even in the color-flavor locked (CFL)
  phase. However, since in the CFL phase all quarks are paired, this
  positive charge is compensated by a negative charge, which turns out
  to be situated in a layer of a few tens of fm below the surface, and
  the total charge of CFL strangelets is zero. We also study the
  surface and curvature contributions to the total energy. Due to the
  strong pairing, the energy as a function of the mass number is very
  well reproduced by a liquid-drop type formula with curvature term.
\end{abstract}
\maketitle

\section{Introduction}
From rather general arguments it is expected that at low temperatures
and high densities quark matter is in a color superconducting
state~\cite{colorsup}. More recently~\cite{Alford97,Rapp97} it has
been suggested that the diquark pairing gaps for quark matter at
densities of several times nuclear matter saturation density could be
of the order of $\sim 100$ MeV. Since this could have important
phenomenological consequences in particular for the interior of
compact stars, this has triggered much work on color superconductivity
in dense quark matter (for reviews, see, e.g.,
Ref.~\cite{reviews}). These investigations of the QCD phase diagram
have revealed a very rich phase structure with many different possible
pairing patterns, depending on external conditions such as, for
instance, electrical neutrality or quark masses. The largest diquark
pairing gaps arise from scalar condensates, leading either to the
two-flavor color superconducting (2SC) phase or to the
color-flavor-locked (CFL) phase~\cite{CFL,weakCFL}. The latter pairing
pattern involves strange ($s$) quarks, in addition to the two light
quark flavors, up ($u$) and down ($d$).

If color superconducting quark matter exists in nature, the most
likely place to find it is the interior of compact stars because
matter is compressed there to densities much higher than nuclear
matter saturation density. However, it has been argued that
strange-quark matter (SQM) might be absolutely
stable~\cite{Witten}. Under this hypothesis, even pure strange stars
should exist~\cite{strangestars}, i.e., stars entirely composed of
SQM. Also small lumps of SQM, called ``strangelets,'' might be
stable. Because of their low charge to baryon number ratio $Z/A$,
strangelets have been proposed to populate ultra-high energy cosmic
rays~\cite{cosmicrays}.

In SQM without pairing, the density of strange quarks is supposed to
be smaller than that of light quarks because of their higher mass.
Consequently, SQM and strangelets are positively charged and the
charge neutrality of strange stars has to be achieved via the presence
of electrons. At the surface an atmosphere of electrons
forms~\cite{strangestars} which can potentially be
detected~\cite{Usov,UsovPage} via the emission of electron-positron
pairs from an extremely strong electric field at the surface.

Recently another possible picture of the surface of a strange star has
been proposed~\cite{nuggets}: there could be a ``crust'' composed of
strangelets immersed in an electron gas. Similar to an ordinary neutron
star, there could be an interface between the crust and the interior
in form of the famous ``pasta phases.'' Within this scenario the
electric field at the surface would be strongly reduced. Obviously,
surface effects for the strangelets play an important role for the
description of this scenario. For instance, there is a critical
surface tension deciding whether a homogeneous phase or the droplet
phase is favored~\cite{stablenuggets}.  Another question for which
surface effects should be considered is the formation of a strange
star in a supernova explosion. Before the explosion the original star
contains hadronic matter. During the formation of the star, nucleation
of strangelets sets in, leading then to a conversion of the entire
star to SQM. For the nucleation process the properties of small
strangelets are important.

Pairing tends to reduce the differences in density of different quark
species.  For bulk quark matter in the CFL state, requiring color
neutrality, all quarks are paired. The densities are thus equal and
CFL quark matter is electrically neutral on its own, i.e., without any
electrons~\cite{RW01}. This would suggest dramatic changes in the
properties of strangelets and SQM inside compact stars. For instance,
the electrosphere at the surface of a strange star could completely
disappear. But, the presence of the surface can modify this picture
since it can lead to a non-zero surface charge which remains even for
large objects. For example, the boundary condition of the MIT bag
model suppresses the density of the massive strange quarks at the
surface, resulting in a positive surface
charge~\cite{Madsen01}. Within this scenario, the total charge of a
strangelet, following roughly $Z \approx 0.3 A^{2/3}$, is drastically
reduced with respect to ``normal'' strangelets. For strange stars,
this requires the presence of electrons~\cite{Usov04}. However,
pairing has not been treated self-consistently in previous work (see,
e.g., Ref.~\cite{Madsen01}). In this paper we will therefore
reinvestigate finite-size strangelets with pairing by considering
quark matter in a color superconducting spherical bag, solving the
Hartree-Fock-Bogoliubov (HFB) equations. We will show, in particular,
that there exist CFL type solutions where all quarks are paired and
the total charge of the strangelet strictly vanishes.

The outline of the paper is as follows. In Section~\ref{sec:model} we
will present our model for treating color superconducting quark matter
in a finite volume. In Section~\ref{sec:results} we will show
numerical results. In Section~\ref{sec:solutions} we discuss the
possibility of qualitatively different configurations. In
Section~\ref{sec:chdens} we concentrate on the charge-density
distributions of the CFL like solutions. In Section~\ref{sec:liquiddrop}
we discuss a liquid-drop like mass formua for the CFL-like solutions
and calculate the surface tension. Finally, in Section~\ref{sec:sum}
we will summarize our results.

\section{Model}
\label{sec:model}
\subsection{Lagrangian}
Since it is not possible to describe strangelets or SQM with a surface
from first principles (QCD), we will use a quark model which allows to
describe finite-size objects. For this purpose we will use here the
MIT bag model~\cite{MITbag}. The idea of this model is that
confinement can be simulated by the existence of a ``bag'' which
consists of a ``hole'' in the non-perturbative QCD vacuum. Inside this
``bag'', the vacuum is supposed to be perturbative, i.e., inside the
bag the interactions of the quarks can be treated perturbatively. To
create this ``hole'' in the non-perturbative QCD vacuum, an energy per
volume, $B$, is necessary. In the present work we will consider a static
spherical bag with radius $R$. On the surface of the bag, the quark
field $\psi$ has to satisfy an appropriate boundary condition. In the
simplest version of the MIT bag model, the boundary condition reads
\begin{equation}
-i \vek{e}_r\cdot \vek{\gamma} \psi = \psi|_{r = R}\,.
\label{boundarycondition}
\end{equation}
which ensures that there is no particle flux across the surface. By $r
= |\vek{r}|$ we denote the radial coordinate, measured from the center
of the bag, and $\vek{e}_r = \vek{r}/r$ is the radial unit vector.
The boundary condition (\ref{boundarycondition}) leads to a
suppression of the wave functions of massive particles at the
surface. This means that the strange-quark density will a priori be
suppressed at the surface with respect to the light quark densities.

The MIT bag model can be expressed in terms of a Lagrangian density as
follows~\cite{Bhaduri}:
\begin{multline}
\mathcal{L}_{\mathit{bag}} =
  [\bar{\psi}(i\gamma^\mu\partial_\mu-m)\psi-B]\theta(R-r)\\
    -\frac{1}{2}\bar{\psi}\psi\delta(R-r)\,,
\label{Lbag}
\end{multline}
where $m$ is the matrix of quark masses. Due to the second term, the
boundary condition (\ref{boundarycondition}) follows immediately from
the Euler-Lagrange equation for the quark field \cite{Bhaduri}.

In order to include pairing, we will supplement the bag model with a
pairing interaction. In principle, perturbative one-gluon exchange
generates an attractive paring interaction in certain channels, in
particular in the scalar color antitriplet channel. For simplicity, we
will use here a four-point pairing interaction acting only in this
dominant channel. The corresponding Lagrangian reads (see any of the
standard review articles on color superconductivity \cite{reviews})
\begin{equation}
\mathcal{L}_\mathit{pair} = H \sum_{A,A^\prime} (\bar{\psi} i \gamma_5
  \tau_A \lambda_{A^\prime} C\bar{\psi}^T) (\psi^T C i \gamma_5 \tau_A
  \lambda_{A^\prime} \psi)\,,
\end{equation}
where $H$ is a dimensionful coupling constant, $C$ denotes the charge
conjugation matrix, and $\tau_A, \lambda_{A^\prime}$ represent $SU(3)$
matrices in flavor and color space, respectively. We follow the
convention that capital letters $A, A^\prime$ indicate that we are
restricting $\tau_A$ and $\lambda_{A^\prime}$ to be antisymmetric,
i.e., in terms of the Gell-Mann matrices, $A,A^\prime \in
\{2,5,7\}$.

In addition to the strong interaction, the quarks will exhibit
electromagnetic interactions which, due to their long range, become
particularly important for large objects. The corresponding Lagrangian
reads
\begin{equation}
\mathcal{L}_{\mathit{e.m.}} = -\frac{1}{4}F_{\mu\nu}F^{\mu\nu}
  -e\bar{\psi}Q A_\mu\gamma^\mu\psi,
\end{equation}
where $F_{\mu\nu} = \partial_\mu A_\nu - \partial_\nu A_\mu$ and
$A_\mu$ denote respectively the electromagnetic field strength tensor
and four-potential, and $Q$ is the matrix of quark charges in units of
$e$, $Q_u = 2/3$, $Q_d = Q_s = -1/3$.

It would be in the spirit of the bag model to include also the gluon
exchange in a perturbative way, i.e., in the same way as the
photon. However, this goes beyond the scope of the present paper and
will be postponed to a future study.

\subsection{Solution in the framework of HFB theory}
\label{sec:hfb}
The model described by $\mathcal{L} = \mathcal{L}_{\mathit{bag}} +
\mathcal{L}_{\mathit{pair}} + \mathcal{L}_{\mathit{e.m.}}$ will be
treated in the framework of HFB theory. By minimizing the energy in
mean field approximation (for more details see Appendix~\ref{app:hfb}
and Ref.~\cite{CH} where the ``Dirac-Hartree-Bogoliubov''
approximation was developped for finite nuclei), one obtains the
following HFB equations:
\begin{equation}
\begin{pmatrix} h & \Delta \\ \Delta & -h
  \end{pmatrix} \begin{pmatrix} U_\alpha(\vek{r}) \\ \gamma^0
      V_\alpha(\vek{r}) \end{pmatrix} = \epsilon_\alpha
      \begin{pmatrix} U_\alpha(\vek{r}) \\ \gamma^0 V_\alpha(\vek{r})
      \end{pmatrix}\,.
\label{hfbequations}
\end{equation}
The single-particle Hamiltonian
\begin{equation}
h = -i \vek{\alpha} \cdot\vek{\nabla} + m \gamma^0 + \Sigma - \mu
\end{equation}
includes besides the free Dirac Hamiltonian the quark self-energy
$\Sigma$ (in our case due to Coulomb interaction) and the matrix of
chemical potentials $\mu$ which depend on flavor $f \in \{u,d,s\}$ and
color $c \in \{r,g,b\}$ (we will denote the three colors by red,
green, and blue). $\Delta$ denotes the pairing field (gap). The
spinors $U_\alpha$ and $V_\alpha$ describe the particle- and hole-like
components of the quark fields, respectively [see
Eq.~(\ref{greensfunctions})], where $\alpha$ is a multi-index
containing all quantum numbers characterizing a single-particle state
(see Appendix~\ref{app:basis}). In writing Eq.~(\ref{hfbequations}),
we implicitly assumed that the pairing field $\Delta$ can be chosen
real, which is the case for the pairing pattern we consider, and that
the self-energy $\Sigma$ is local, which is the case since we neglect
the exchange (Fock) term (see below).

The pairing field $\Delta$ and the self-energy $\Sigma$ depend
themselves on the wave functions $U$ and $V$, such that we have to
solve a self-consistency problem. To be specific, the pairing field
$\Delta$ depends on the diquark condensates
\begin{equation}
s_{AA^\prime}(x) = - \langle \bar{\psi_T}(x) \tau_A \lambda_{A^\prime}
  \psi(x) \rangle\,,
\label{saa}
\end{equation}
where $\psi_T$ denotes the time-reversed conjugate of
$\psi$,
\begin{equation}
\psi_T = \gamma_5 C \bar{\psi}^T\,.
\end{equation}
The diquark condensates can be expressed in terms of the $U$ and $V$
functions as
\begin{equation}
s_{AA^\prime}(r) = -\sum_{\beta, \epsilon_\beta < 0} 
  \bar{V}_\beta(\vek{r}) \tau_A\lambda_{A^\prime} U_\beta(\vek{r})
\label{saadivergent}
\end{equation}
(since we are dealing with a static problem, the condensates do not
depend on time, and due to spherical symmetry, they depend only on the
radial coordinate $r$). We will limit our investigations here to
diagonal condensates, i.e., only condensates with $A = A^\prime$ are
non-zero\footnote{In uniform infinite matter it can be shown
\cite{CFL} that for the energetically favored solution the arbitrary
orientation in color can be chosen in such a way that only the
diagonal condensates with $A = A^\prime$ are non-zero.}. In uniform
infinite matter and for an exact $SU(3)$ flavor symmetry, the CFL
phase is characterized by nonzero values $s_{22} = s_{55} = s_{77}$,
whereas the 2SC state has only $s_{22} \neq 0$. The relation between
the condensates $s_{AA}$ and the pairing field $\Delta$ reads
\begin{gather}
\Delta(r) = \sum_{A=2,5,7} \Delta_A(r) \tau_A \lambda_A\,,\\
\Delta_A(r) = 2\, H s_{AA}(r)\,.
\label{gapequation}
\end{gather}
In practice, the expression (\ref{saadivergent}) is divergent and it
is necessary to introduce a cutoff in order to obtain a finite
result. Since in a finite system the levels are discrete, a sharp
cutoff would generate discontinuities as a function of the system's
size. We therefore introduce a smooth cutoff function $f(p/\Lambda)$
(see Appendix~\ref{app:cutoff} for details). Another practical problem
arises from antiparticle contributions. However, since the chemical
potentials $\mu_{fc}$ are large and positive and pairing involves
mostly the states near the Fermi surface, we assume that the
antiparticle contributions are not important and can be neglected. We
checked this approximation (analogous to the ``no-sea approximation''
in nuclear physics~\cite{CH}) in infinite matter and found that the
effect of antiparticle states can be absorbed in a readjustment of the
coupling constant by $\sim 20\,\%$.

For the normal self-energy $\Sigma$ we employ the Hartree
approximation, i.e., we neglect the Coulomb exchange (Fock) term as
well as exchange contributions from the magnetic field. We also
disregard the contribution of $\mathcal{L}_{\mathit{pair}}$ to the
normal self-energy. Hence, the self-energy is simply proportional to
the static Coulomb potential
\begin{equation}
\Sigma(r) = e Q A_0(r)\gamma^0\,.
\end{equation}
The Coulomb potential is related to the quark densities by
\begin{equation}
A_0(r) = e \int d^3 r^\prime
  \frac{\rho_\mathit{ch}(\vek{r}^\prime)}{|\vek{r}-\vek{r}^\prime|}\,,
\label{vcoulomb}
\end{equation}
where
\begin{equation}
\rho_\mathit{ch}(r) = \sum_f Q_f\rho_f(r)
\end{equation}
is the charge density (divided by $e$), $\rho_f$ being the number
density of quarks of flavor $f$. As it was the case for the diquark
condensates, the quark number densities can be expressed in terms of
the $U$ and $V$ functions. Denoting by $\tilde{\beta}$ all
single-particle quantum numbers except flavor, we can write the number
density of quarks of flavor $f$ as
\begin{equation}
\rho_f(r) = \sum_{\tilde{\beta}, \epsilon_{f\tilde{\beta}} < 0}
  U^\dagger_{f\tilde{\beta}}(\vek{r}) U_{f\tilde{\beta}}(\vek{r})\,.
\label{rhof}
\end{equation}

Let us now summarize the procedure how the HFB equations are
solved. We start with an initial guess for the pairing fields
$\Delta_A(r)$ and for the Coulomb potential $A_0(r)$. Then we solve
the eigenvalue problem (\ref{hfbequations}) in order to find the $U$
and $V$ functions. From these functions the diquark condensates
$s_{AA}(r)$ and the quark densities $\rho_f(r)$ are computed according
to Eqs.~(\ref{saadivergent}) and (\ref{rhof}), which are then used to
update the pairing fields $\Delta_A$ and the Coulomb field $A_0$
according to Eqs.~(\ref{gapequation}) and (\ref{vcoulomb}). These
steps are iterated until convergence (i.e., self-consistency) is
reached.

The crucial difference to the BCS formalism in homogeneous infinite
matter is that in our case the wave functions adapt themselves to the
pairing field and to the Coulomb potential, whereas in the case of
homogeneous infinite matter the wave functions always stay plane
waves, and the $U$ and $V$ factors are just coefficients multiplying
them.
\subsection{Determination of chemical potentials and bag radius}
In Section~\ref{sec:hfb} we described how the HFB equations are
solved for given values of the chemical potentials $\mu_{fc}$ and of
the bag radius $R$. However, in reality, only one quantity is given,
namely the baryon number $A$. Even the fractions of different quark
flavors cannot be fixed, unless one allows for $\beta$ unstable
strangelets. Let us now describe how we determine the chemical
potentials $\mu_{fc}$ and the bag radius $R$ for given baryon number
$A$.

The first step consists in fixing the quark numbers, $N_{fc}$, for
each flavor $f$ and color $c$, and to adjust the chemical potentials
$\mu_{fc}$ in order to obtain these quark numbers. Before we address
the question how the nine quark numbers $N_{fc}$ are determined, let
us discuss the issue of the bag radius $R$. Until now, the radius was
imposed from outside, but in reality the system will choose its radius
such that it minimizes its total energy for given quark numbers
$N_{fc}$. Within the bag model, the total energy is given by
\begin{equation}
E = E_q+BV\,,
\label{Etotal}
\end{equation}
where $V = 4\pi R^3/3$ is the volume of the bag. By $E_q$ we denote
the energy of the quarks inside the bag, including the interaction
energy, which in our case comes from pairing and Coulomb
interactions. It can be obtained from the solution of the HFB
equations as follows \cite{CH}:
\begin{multline}
E_q = \int_{r<R} d^3 r \sum_{\beta,\epsilon_\beta < 0}
  \left(U^\dagger_\beta(\vek{r})
    (\epsilon_\beta+\mu)U_\beta(\vek{r})\phantom{\frac{1}{2}}\right.\\
  \left.+\frac{1}{2}\left[\bar{U}_\beta(\vek{r})\Delta(r)V_\beta(\vek{r})
  -U^\dagger_\beta(\vek{r})e Q
  A_0(r)U_\beta(\vek{r})\right]\right)\,.
\label{quarkenergy}
\end{multline}
Minimizing the total energy $E$ is of course completely equivalent to
saying that the quark pressure in the bag is counterbalanced by the
bag pressure $B$, i.e.,
\begin{equation}
\left.\frac{dE_q}{dV}\right|_N = -B\,.
\label{dEdV}
\end{equation}
This equation determines the radius of the strangelet for given bag
pressure $B$, interaction strength $H$ and quark numbers $N_{fc}$. In
practice, however, we find it more convenient to minimize $E$ rather
than solve Eq.~(\ref{dEdV}).

Let us now turn to the determination of the quark numbers. The nine
quark numbers $N_{fc}$ cannot be chosen arbitrarily, but they have
to fulfil certain requirements. Imposing the total baryon number $A$
and color neutrality, i.e., equal numbers of quarks for each color, we
have to satisfy the constraint
\begin{equation}
\sum_f N_{fc} = A \qquad \mbox{for all $c$}\,.
\label{numberconstraint}
\end{equation}
Of course, these three equations are not sufficient for determining
all the nine quark numbers. In order to get unique values for the
$N_{fc}$, it is necessary to impose $\beta$ stability, as we will
describe now.

In an infinite homogeneous system the condition for $\beta$
equilibrium gives just a relation between the chemical
potentials\footnote{Here we assume that neutrinos are not trapped,
i.e., they can freely leave the system}
\begin{equation}
\mu_{dc} = \mu_{sc} = \mu_{uc} + \mu_{e}\qquad\mbox{for all $c$}\,.
\label{betainfinite}
\end{equation}
In a small system this is slightly different. First, even if there are
electrons (i.e., if the strangelet is charged), they are not localized
inside the strangelet, but they form a large cloud like in ordinary
atoms and hence their chemical potential $\mu_e$ is approximately
equal to the electron mass and can be neglected. Without pairing, it
has been estimated in Ref.~\cite{FarhiJaffe} that this may be still
the case for strangelets with charge $Z \lesssim 1000$, corresponding
roughly to $A\lesssim 10^6$. The second difference to bulk matter
comes from the fact that, due to the discrete levels, particle numbers
are discontinuous functions of the chemical potentials. The term
$\beta$ equilibrium should now be replaced by $\beta$ stability, which
means that the system does not gain energy by performing a $\beta$
decay, inverse $\beta$ decay, or electron capture, i.e., transforming
an up into a down or strange quark, or vice versa, accompanied by the
corresponding leptons.

To achieve $\beta$ stability, we therefore compare the energies of
adjacent strangelets with the same total quark number per color,
differing only in the number of up, down, and strange quarks,
respectively, in order to find the configuration with the lowest total
energy $E$. Of course, in the case of large particle numbers, the
minimum-energy configuration fulfils approximately the
condition~(\ref{betainfinite}).

\subsection{Choice of the model parameters}
Besides the quark masses, which we take as $m_u = m_d = 0$ and $m_s =
120$ MeV, our model contains three parameters: the bag constant $B$,
the coupling constant of the pairing interaction, $H$, and the cutoff
$\Lambda$ which is necessary to avoid the divergence of the gap
equation (\ref{saadivergent}), see below Eq.~(\ref{gapequation}). In
fact, a change of the cutoff in reasonable limits can to very good
approximation be compensated by a change of the coupling constant. We
therefore choose rather arbitrarily $\Lambda = 600$ MeV and give
instead of the dimensionful coupling constant $H$ the dimensionless
combination $H\Lambda^2$. So we are left with two parameters, $B$ and
$H\Lambda^2$

We can get an idea of the value of the bag pressure by looking at the
stability of bulk quark matter. Non-strange quark matter should be
energetically less favored than normal hadronic matter, whereas SQM
should be stable if for some baryon number $A > A_c$ strangelets
become stable and consequently strange stars can exist. This means
that we want the energy per baryon of SQM to be less than 931 MeV, the
energy per baryon of the most stable nucleus, $^{57}$Fe. On the other
hand, the energy per baryon of non-strange quark matter should be
larger than the nucleon mass. Without interaction the window for the
values of the bag constant is then $148$ MeV $< B^{1/4} < 157$
MeV. These values change as a function of the interaction strength
$H$. To better compare the results, we will readjust for each coupling
strength the bag constant in order to get $E/A = 900$ MeV. The
corresponding values are listed in Table~\ref{tab:bagvalue}, together
with other properties of infinite matter. Non-strange quark matter is
unstable with these parameter values. Note that for the weakest
non-vanishing coupling constant given in Table~\ref{tab:bagvalue}, SQM
is in the 2SC phase and not in the CFL phase.
\begin{table}
\caption{\label{tab:bagvalue} Values of the bag constants for
different values of the coupling constant $H$, resulting in color and
electrically neutral SQM with electrons in $\beta$ equilibrium with an
energy per baryon of $E/A = 900$ MeV. The corresponding baryon
densities $\rho_B$, electron densities $\rho_e$, and pairing gaps in
infinite matter are also displayed.}
\begin{ruledtabular}
\begin{tabular}{ccccccc}
$H \Lambda^2$ &
$\begin{matrix}B^{1/4}\\ \mbox{(MeV)}\end{matrix}$ &
$\begin{matrix}\rho_B\\ \mbox{(fm$^{-3}$)}\end{matrix}$ &
$\begin{matrix}\rho_e\\ \mbox{(fm$^{-3}$)}\end{matrix}$ &
$\begin{matrix}\Delta_2\\ \mbox{(MeV)}\end{matrix}$ &
$\begin{matrix}\Delta_5=\Delta_7\\ \mbox{(MeV)}\end{matrix}$\\
\hline
0    & 152.03 & 0.329 & 7.3$\times 10^{-6}$ & 0    & 0   \\
1.5  & 152.44 & 0.339 & 9.7$\times 10^{-5}$ & 27.7 & 0   \\
1.75 & 153.97 & 0.367 & 0                   & 35.1 & 34.5\\
2    & 156.26 & 0.395 & 0                   & 50.6 & 49.7\\
2.25 & 159.46 & 0.427 & 0                   & 67.2 & 66.0\\
2.5  & 163.46 & 0.463 & 0                   & 84.6 & 83.1
\end{tabular}
\end{ruledtabular}
\end{table}
For the larger coupling constants, the CFL phase is preferred. Note
that, due to the mass difference of light and strange quarks, the
flavor $SU(3)$ symmetry is not exact and the gap $\Delta_2$ is
different from $\Delta_5$ and $\Delta_7$. However, since the CFL phase
is electrically neutral, and we have $m_u= m_d = 0$, the isospin
$SU(2)$ symmetry in the up- and down-quark sector is exact and
therefore $\Delta_5 = \Delta_7$.
\section{Results}
\label{sec:results}
\subsection{Different types of solutions}
\label{sec:solutions}
\begin{figure}
\includegraphics[width=7cm]{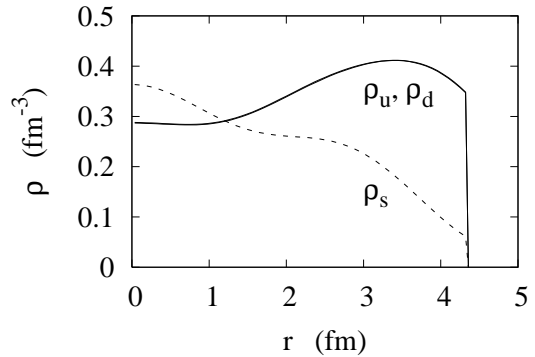}
\caption{Quark number density profiles of the strangelet $A = 108$, $Z
= 24$ in the case of vanishing pairing interaction (free quarks in a
bag) and $B^{1/4} = 152.03$ MeV.}
\label{fig:rho_H0_A108_Z24}
\end{figure}

We will first discuss the qualitatively different configurations we
find. Let us start by discussing a small strangelet ($A = 108$, $Z =
24$) without any pairing interaction ($H\Lambda^2 = 0$). The mass
number has been chosen such that the minimum-energy configuration is a
closed-shell configuration.  The quark numbers and other relevant
information are listed in Table~\ref{tab:strangelets}.
\begin{table*}
\caption{\label{tab:strangelets} Parameters and properties of the
strangelets discussed in Section~\ref{sec:solutions}: $B = $ bag
constant, $H = $ coupling constant of the pairing interaction, $A = $
baryon number, $Z = $ charge, $N_{fc} =$ number of quarks of flavor
$f$ and color $c$, $E/A = $ energy per baryon, $R = $ radius of the
bag, $\Delta_A(0)$ = value of the gap at $r = 0$.}
\begin{ruledtabular}
\begin{tabular}{cccccccccc}
$\begin{matrix}B^{1/4}\\\mbox{(MeV)}\end{matrix}$ &
$\begin{matrix}H\Lambda^2\\ \mbox{(MeV)}\end{matrix}$ &
$A$ &
$Z$ &
\rule[-3ex]{0pt}{7ex} $\left(\begin{smallmatrix}
                N_{ur}& N_{ug}& N_{ub}\\
                N_{dr}& N_{dg}& N_{db}\\
                N_{sr}& N_{sg}& N_{sb}\end{smallmatrix}\right)$ &
$\begin{matrix} E/A\\ \mbox{(MeV)}\end{matrix}$ &
$\begin{matrix} R\\ \mbox{(fm)}\end{matrix}$ &
$\begin{matrix} \Delta_2(0)\\ \mbox{(MeV)}\end{matrix}$ &
$\begin{matrix} \Delta_5(0)\\ \mbox{(MeV)}\end{matrix}$ &
$\begin{matrix} \Delta_7(0)\\ \mbox{(MeV)}\end{matrix}$
\\
\hline
152.03 & 0 & 108 & 24 &
\rule[-2.1ex]{0pt}{5.5ex} $\left(\begin{smallmatrix}
  44&44&44\\44&44&44\\20&20&20\end{smallmatrix}\right)$ &
932.5 & 4.36 & 0 & 0 & 0
\\
152.44 & 1.5 & 108 & 24 &
\rule[-2.1ex]{0pt}{5.5ex} $\left(\begin{smallmatrix}
  44&44&44\\44&44&44\\20&20&20\end{smallmatrix}\right)$ &
930.7 & 4.31 & 32.9 & 0 & 0
\\
153.97 & 1.75 & 108 & 24 &
\rule[-2.1ex]{0pt}{5.5ex} $\left(\begin{smallmatrix}
  44&44&44\\44&44&44\\20&20&20\end{smallmatrix}\right)$ &
934.0 & 4.24 & 49.5 & 0 & 0
\\
153.97 & 1.75 & 108 & 10 &
\rule[-2.1ex]{0pt}{5.5ex} $\left(\begin{smallmatrix}
  38&39&41\\39&38&41\\31&31&26\end{smallmatrix}\right)$ &
934.8 & 4.21 & 41.6 & 24.9 & 24.9
\\
153.97 & 1.75 & 108 & 0 &
\rule[-2.1ex]{0pt}{5.5ex} $\left(\begin{smallmatrix}
  38&35&35\\35&38&35\\35&35&38\end{smallmatrix}\right)$ &
934.8 & 4.17 & 33.9 & 37.0 & 36.9
\end{tabular}
\end{ruledtabular}
\end{table*}
Due to the finite size of the bag, the energy per baryon
($E/A = 932.5$ MeV, including 1.0 MeV due to Coulomb) is much higher
than that of color neutral infinite matter with $\mu_{uc} = \mu_{dc} =
\mu_{sc}$\footnote{As discussed below Eq.~(\ref{betainfinite}), it is
more appropriate to compare a small strangelet with this kind of
matter rather than electrically neutral matter with electrons in $\beta$
equilibrium.}  ($E/A = 899.5$ MeV). This effect will be discussed in
more detail in Section~\ref{sec:liquiddrop}. The density profiles of
light and strange quarks are shown in Fig.~\ref{fig:rho_H0_A108_Z24}.
As expected, due to the boundary condition, the strange-quark density
is strongly suppressed at the surface, contrary to the densities of
the light quarks. For comparison we mention that for the same value of
the bag constant, the densities in color neutral infinite matter with
$\mu_{uc} = \mu_{dc} = \mu_{sc}$ are: $\rho_u = \rho_d = 0.355$
fm$^{-3}$, $\rho_s = 0.274$ fm$^{-3}$. We see that not only the
strange-quark density, but also the densities of the light quarks are
quite different from these values and depend strongly on $r$ because
of the existence of discrete levels in the bag. Let us mention that,
due to the Coulomb potential, the density profiles of up and down
quarks are slightly different, but the difference is too small to be
visible in Fig.~\ref{fig:rho_H0_A108_Z24}.

\begin{figure}
\includegraphics[width=7cm]{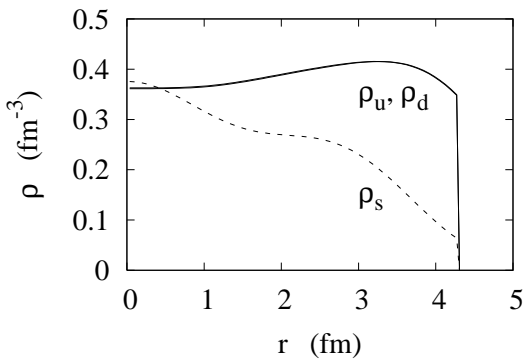}
\caption{Quark number density profiles of the strangelet $A = 108$, $Z
= 24$ in the case of $H\Lambda^2 = 1.5$ and $B^{1/4} = 152.44$ MeV.}
\label{fig:rho_H1.5_A108_Z24}
\end{figure}

\begin{figure}
\includegraphics[width=7cm]{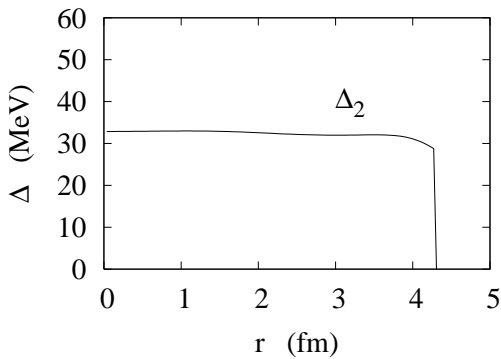}
\caption{Gap $\Delta_2(r)$ of the strangelet $A = 108$, $Z
= 24$ in the case of $H\Lambda^2 = 1.5$ and $B^{1/4} = 152.44$ MeV.}
\label{fig:delta_H1.5_A108_Z24}
\end{figure}

Now we switch on the pairing interaction. In the case of $H\Lambda^2 =
1.5$, SQM is in the 2SC phase, i.e., only up and down quarks of two
colors (red and green in our notation) are paired. This is also true
in a finite strangelet. Therefore it is clear that the strange-quark
density profile remains the same as without pairing. The oscillations
of the densities of the light quarks, however, are much weaker now
than in the case without pairing, since pairing washes out the
occupation numbers. This can be seen in
Fig.~\ref{fig:rho_H1.5_A108_Z24}.
In this 2SC-like solution, only one of the gaps, $\Delta_2$, is
non-zero. Since $\Delta_2$ involves only the wave functions of up and
down quarks, which are not suppressed at the surface, it extends up to
the surface of the bag, as shown in
Fig.~\ref{fig:delta_H1.5_A108_Z24}.
As a function of $r$, it is almost constant and quite close to the
corresponding value in infinite matter with $\mu_{uc} = \mu_{dc} =
\mu_{sc}$, which is $\Delta_2 = 29.2$ MeV.

If we increase the coupling constant to $H\Lambda^2 = 1.75$, we obtain
three qualitatively different solutions which have comparable
energies. The most stable one is still of the 2SC type, although in
infinite matter the CFL phase is preferred. In this case, the
strangelet still has $Z = 24$ and the density profiles are almost
identical to those shown in Fig.~\ref{fig:rho_H1.5_A108_Z24}. The main
difference is that now the value of the gap is larger.

\begin{figure}
\includegraphics[width=7cm]{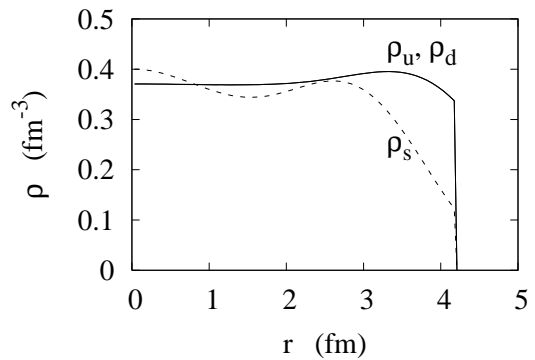}
\caption{Density profiles of the strangelet $A = 108$, $Z
= 10$ in the case of $H\Lambda^2 = 1.75$ and $B^{1/4} = 153.97$ MeV.}
\label{fig:rho_H1.75_A108_Z10}
\end{figure}

\begin{figure}
\includegraphics[width=7cm]{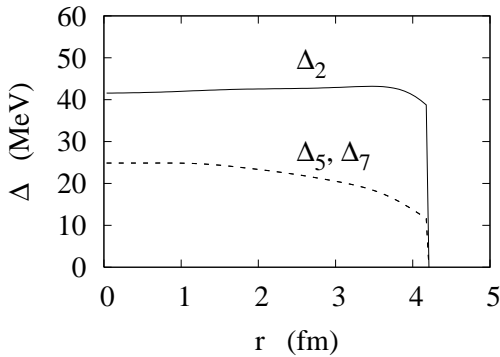}
\caption{Gaps $\Delta_A$ as functions of $r$ for the strangelet $A =
108$, $Z = 10$ in the case of $H\Lambda^2 = 1.75$ and $B^{1/4} =
153.97$ MeV.}
\label{fig:delta_H1.75_A108_Z10}
\end{figure}

In the two other solutions, also strange quarks participate in pairing
($\Delta_5 \approx \Delta_7\neq 0$ -- note that $\Delta_5$ and
$\Delta_7$ are not exactly equal because the isospin symmetry is
broken by the Coulomb interaction). These two solutions have charge $Z
= 10$ and $Z = 0$, respectively. Let us first discuss the case $Z =
10$. In this case, there are a couple of up and down quarks which
remain unpaired. The wave function of the unpaired level is mainly
localized near the surface of the bag, as can be seen in
Fig.~\ref{fig:rho_H1.75_A108_Z10}, where the density profiles are
shown.
In the inner part, the densities of up, down, and strange quarks are
almost equal, while near the surface, where the strange-quark density
is suppressed due to the boundary condition, there is an excess of up
and down quarks. This excess is due to the unpaired quarks. The fact
that one level of up and down quarks (in the present case the
$1g_{9/2}$ level, i.e., the lowest level with $j = 9/2, \kappa = -5$
in the notation of Appendix~\ref{app:basis}) does not participate in
pairing means that the occupation number of this level is equal to
1. At the same time, the corresponding level of the strange quarks has
an occupation number equal to 0. In a certain sense this situation is
analogous to the ``breached pairing'' phase of infinite
matter~\cite{LiuWilczek}. The charge $Z$ is equal to the degeneracy
$2j+1$ of the unpaired level. The gaps $\Delta_A$ as functions of $r$
corresponding to this solution are displayed in
Fig.~\ref{fig:delta_H1.75_A108_Z10}.

\begin{figure}
\includegraphics[width=7cm]{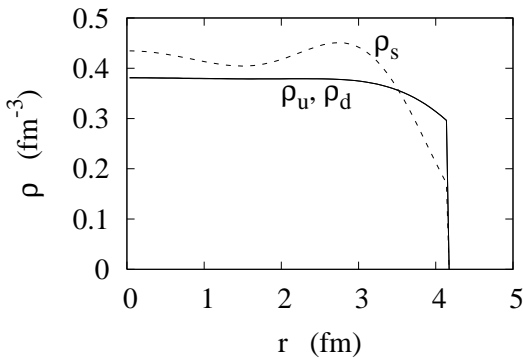}
\caption{Density profiles of the strangelet $A = 108$, $Z
= 0$ in the case of $H\Lambda^2 = 1.75$ and $B^{1/4} = 153.97$ MeV.}
\label{fig:rho_H1.75_A108_Z0}
\end{figure}

\begin{figure}
\includegraphics[width=7cm]{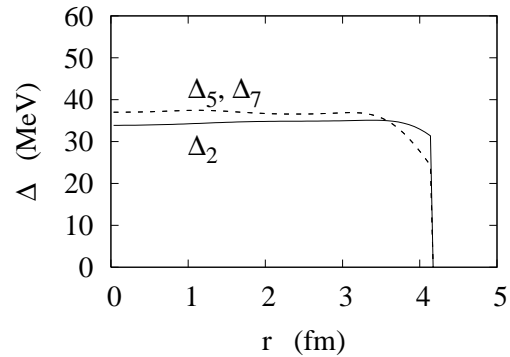}
\caption{Gaps $\Delta_A$ as functions of $r$ for the strangelet $A =
108$, $Z = 0$ in the case of $H\Lambda^2 = 1.75$ and $B^{1/4} =
153.97$ MeV.}
\label{fig:delta_H1.75_A108_Z0}
\end{figure}

In the third solution, all quarks are paired. As a consequence, the
numbers of up, down, and strange quarks are equal, and the total
charge is $Z = 0$. This is analogous to the CFL phase in the infinite
system. Since the strange-quark density is suppressed near the
surface, but the number of strange quarks is equal to that of up and
down quarks, it is clear that the strange-quark density must be larger
than the up- and down-quark densities in some other part of the
system. This is indeed the case, as can be seen in
Fig.~\ref{fig:rho_H1.75_A108_Z0}.
We also see that the excess of the light-quark densities over the
strange-quark density is reduced as compared with the case $Z=10$
discussed above (cf. Fig.~\ref{fig:rho_H1.75_A108_Z10}). We will
discuss the charge-density distribution in detail in
Section~\ref{sec:chdens}. The gaps, shown in
Fig.~\ref{fig:delta_H1.75_A108_Z0}, are much closer to the gaps in
infinite matter (cf. Table~\ref{tab:bagvalue}) than in the case
$Z=10$.

For the larger values of the coupling constant we considered
($H\Lambda^2 = 2, 2.25, 2.5$), it is always the CFL-type solution ($Z =
0$) which has the lowest energy. We do not show any figures because in
all these cases the results are analogous to those shown in
Figs.~\ref{fig:rho_H1.75_A108_Z0} and \ref{fig:delta_H1.75_A108_Z0}
(just the values of the gaps change, they are close to those given in
Table~\ref{tab:bagvalue} for infinite matter).

It should be mentioned that the fully paired solutions with $Z = 0$
are very robust as soon as the coupling constant is sufficiently
large, i.e., we find this type of solution for arbitrary numbers of
quarks\footnote{If the number of quarks is odd, it it impossible to
pair all quarks and one or several state(s) should be ``blocked'' by
the unpaired quark(s). At present, we have not included this effect in
our calculation, and we restrict ourselves to even quark
numbers}. This solution is in contrast to previous findings (see,
e.g., Ref.~\cite{Madsen01}), where it was supposed that the CFL matter
should be neutral in the bulk with just a thin positively charged
surface layer with an excess of up and down quarks because of the
boundary condition. In fact, this idea corresponds roughly to our
solution with unpaired up and down quarks near the surface. This
solution is, however, very fragile and exists only for certain values
of parameters and mass numbers, since it requires the existence of a
suitable level of light and strange quarks near the respective Fermi
surfaces which can serve as unpaired level. 

\subsection{Charge density distribution}
\label{sec:chdens}
We have seen in Section~\ref{sec:solutions} that in all cases except
the 2SC phase, pairing drastically reduces the total charge
$Z$. Because of surface effects, the local charge density does,
however, not vanish, even within the CFL-type solution which has
$Z=0$. Due to the suppression of the strange-quark wave function at
the surface, a positively charged surface layer remains with an
extension of $\sim 1$ fm, as has already been pointed out in
Ref.~\cite{Madsen01}.

Within the configuration with some unpaired light quarks at the
surface, the total charge of the strangelet results from this positive
surface charge, the interior of the strangelet has almost zero
charge density. The total charge is here reduced compared with a
strangelet without pairing, for example the $A = 108$ strangelet has
$Z = 10$ within this paired configuration, whereas the corresponding
unpaired strangelet has $Z = 24$. A systematic study of the total
charge of strangelets in this configuration will not be discussed here
since this configuration is rather fragile with respect to the details
of the single-particle spectra and thus difficult to realize for many
different particle numbers.

\begin{figure}
\includegraphics[width=7cm]{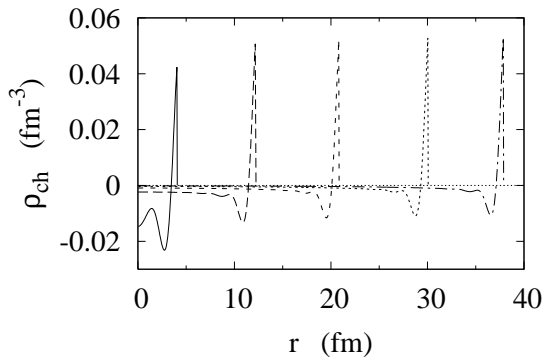}
\caption{Charge density profiles of the fully paired ($H\Lambda^2 =
2$) strangelets $A = $108, 3000, 15000, 45000, and 90000 (from left to
right).}
\label{fig:coulomb1}
\end{figure}

Let us therefore concentrate on the CFL-type solution, which exists
for arbitrary particle numbers. We consider different mass numbers $A$
from $A = 108$ to $A = 90000$, for one particular value of the
coupling constant, $H\Lambda^2 = 2$. In order to reduce the
considerable numerical effort, we use for the large strangelets
(starting from $A = 15000$) the condition (\ref{betainfinite}) with
$\mu_e = 0$ (as a consequence, the quark numbers for each flavor and
color are not integers) instead of looking for the true energy minimum
with respect to $\beta$ decay. In addition, we do not minimize the
energy with respect to the radius, but we simply estimate the volume
of the bag by dividing the mass number $A$ by the baryon density
$\rho_{B\,\mathit{bulk}}$ of infinite matter. These two approximations
are very accurate for such large strangelets. Already in the case of
$A = 3000$, the quark numbers and the radius are very well reproduced
within these approximations: the full minimization results in quark
numbers $N_{ur} = N_{dg} = N_{sb} = 1052$, $N_{ug} = N_{dr} = N_{ub} =
N_{sr} = N_{db} = N_{sg} = 974$, and a radius $R = 12.23$ fm, while
the approximations lead to $N_{ur} = 1051.8$, $N_{dg} = 1051.7$,
$N_{sb} = 1051.1$, $N_{ug} = N_{dr} = 973.8$, $N_{ub} = N_{sr} =
974.4$, $N_{db} = N_{sg} = 974.5$, and $R = 12.19$ fm. Our results for
the charge densities for $A = $108, 3000, 15000, 45000, and 90000 are
shown in Fig.~\ref{fig:coulomb1}.

Since all quarks are paired, we have equal numbers of up, down, and
strange quarks such that the total charge of these strangelets is
zero. The positive surface charge is mostly compensated by an excess
of negative charge concentrated at around 1-3 fm below the surface. We
stress that this concentration of negative charge in a thin layer is a
consequence of pairing and the effect persists if Coulomb interaction
is switched off. In fact, since also the strange quarks are paired,
the ``missing'' strange-quark density at the surface must be
compensated by an ``overshooting'' of the strange-quark density within
a distance corresponding to the size of the Cooper pairs, i.e., the
coherence length $\xi$. Due to the strong gap, the coherence length is
very small: Using the estimate $\xi \sim 1/(\pi\Delta)$, one finds that
it is of the same order as the Fermi wavelength and, strictly
speaking, one might therefore question that mean-field results are
quantitatively correct \cite{Abuki}. The smallness of $\xi$ explains
why the compensation of the negative surface charge is mostly
concentrated in such a thin layer at a small distance from the
surface.

Below this strongly negatively charged layer, the charge density stays
negative but much smaller. Due to Coulomb interaction, which tries to
push the charge towards the surface, this negative charge density
decreases with increasing distance from the surface, especially for
large strangelets. Actually, if Coulomb interaction is switched off,
the remaining charge is distributed more or less homogeneously over
the whole volume.

The behaviour of the charge density far away from the surface in the
presence of Coulomb interaction can easily be interpreted in terms of
Debye screening (similar considerations can be found in
Ref.~\cite{Tatsumi} for the case of hadron-quark mixed phases): We
know that in a uniform medium with Debye screening the Laplace
equation for the Coulomb potential is replaced by
\begin{equation}
\left(\vek{\nabla}^2-\frac{1}{\lambda^2}\right)A_0 = 0\,,
\label{poissonscreened}
\end{equation}
where $\lambda$ is the screening length, which can be obtained from
the limit $\Pi^{00}(q^0 = 0,\vek{q}\to 0)$, where $\Pi^{\mu\nu}(q)$ is
the polarization tensor in the uniform system. This is equivalent to
the expression \cite{Tatsumi}
\begin{equation}
\frac{1}{\lambda^2} = 4\pi e^2 \sum_{fc} Q_f 
  \frac{\partial\rho_{\mathit{ch}}}{\partial \mu_{fc}}\,.
\end{equation}
Computing numerically this derivative within our model for the case of
bulk CFL matter with $B = 156.26$ MeV and $H\Lambda^2 = 2$, we obtain
$\lambda = 7.74$ fm.

\begin{figure}
\includegraphics[width=7cm]{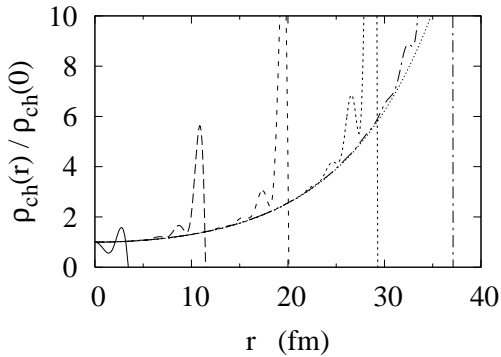}
\caption{Zoom into the part of Fig.~\ref{fig:coulomb1} where the
charge density behaves as given by Eq.~(\ref{chargecenter}). For a
better visibility, the charge densities have been divided by their
respective values at $r = 0$. The thin dotted curve corresponds to
Eq.~(\ref{chargecenter}) with $\lambda = 7.74$ fm.}
\label{fig:coulomb2}
\end{figure}

Taking the Laplacian of Eq.~(\ref{poissonscreened}), we see that the
charge density obeys the analogous equation
\begin{equation}
\left(\vek{\nabla}^2-\frac{1}{\lambda^2}\right)\rho_{\mathit{ch}} =
0\,.
\end{equation}
In the case of half-infinite matter with a surface at $z = 0$, the
solution of this equation shows that the charge density goes to zero
as $\rho_{\mathit{ch}}\propto\exp(z/\lambda)$ if one goes away from
the surface ($z\to -\infty$). In the case of a sphere, the
corresponding solution reads
\begin{equation}
\rho_{\mathit{ch}} \propto \frac{\sinh(r/\lambda)}{r/\lambda}\,.
\label{chargecenter}
\end{equation}
Far away from the surface, the charge densities which we obtain are
very well described by Eq.~(\ref{chargecenter}). To show this, we
display in Fig.~\ref{fig:coulomb2} the same charge densities as in
Fig.~\ref{fig:coulomb1}, but divided by their value at $r = 0$. Far
away from the surface, all curves follow exactly
Eq.~(\ref{chargecenter}) with the value $\lambda = 7.74$ fm calculated
for bulk CFL matter. Near the surface, i.e., at distances which are of
the order of a couple of Fermi wavelengths, there are strong
deviations from this behavior due to Friedel-type oscillations
\cite{GarciaMoliner}. This is because Eq.~(\ref{poissonscreened}) is
not exact, but it is only valid in a uniform medium and in the
long-wavelength limit.

It is interesting to notice that the value of the Debye screeing
length we obtain is in reasonable agreement with the photon Debye mass
calculated from perturbative QCD, which reads, for the CFL phase,
$m_{D,\gamma\gamma}^2 = 1/\lambda^2 = 4 \frac{21-8\ln 2}{54} e^2 N_f
\mu^2/(6 \pi^2)$~\cite{SWR04} ($m_{D,\gamma\gamma}$ denotes the Debye
mass without gluon-photon mixing, see below). For typical values of
the chemical potential this gives $\lambda \sim 10$ fm.

In principle, in color superconducting phases, the photon can mix with
one of the gluons. In the CFL phase, in bulk matter, one linear
combination of photon and gluon stays massless. This means that at
large distances $d\gg \xi$, the Debye screening for the ``rotated''
photon~\cite{ABR00,LitimManuel} does not work, since the Cooper pairs
are neutral with respect to the rotated charge $\tilde{Q}$. Within the
simple model we use for the moment, there are no gluons, such that the
mixing cannot be studied. It could be taken into account, as mentioned
at the end of Section~\ref{sec:model}, by including the gluons in the
same way as the photon, i.e., on the Hartree level. We expect that if
we included the gluons in this way, we would find an even faster
decrease of the charge if we go away from the surface, since in
addition to the electromagnetic force we would have the color forces,
which try to push the color charges to the surface, and in the CFL
phase color neutrality goes hand in hand with electrical
neutrality. Therefore, this is not in contradiction with the fact that
the rotated photon is massless, but it is just a consequence of the
fact that the combination of photon and gluon which is orthogonal to
the rotated photon is massive (in fact, it is even heavier than the
other gluons~\cite{LitimManuel}). This means that in a large object,
like a strange star, all the negative charge will be concentrated
within a layer of a thickness of at most a few tens of fm below the
surface.  However, before drawing any firm conclusion, one should
study this problem in more detail. This will be left for future work.
\subsection{Liquid-drop type expansion}
\label{sec:liquiddrop}
The advantage of the present approach is that finite size effects are
correctly implemented. For large numbers of particles, this becomes,
however, rather cumbersome and asymptotic expansions such as a
liquid-drop type approach can be very useful. We will discuss here the
determination of the parameters, such as the surface tension, of a
liquid-drop type formula for the energy per baryon as a function of
the baryon number $A$, including a surface and a curvature term,
\begin{equation}
\frac{E}{A} = \left(\frac{E}{A}\right)_{\mathit{bulk}}
  + \frac{a_S}{A^{1/3}} + \frac{a_C}{A^{2/3}}\,,
\label{liquiddrop}
\end{equation} 
from our results. As in Section~\ref{sec:chdens}, we will restrict our
discussion to the CFL-type solutions with $Z = 0$, such that we do not
need to include a Coulomb term $\propto Z/A^{1/3}$.

\begin{figure}
\includegraphics[width=7cm]{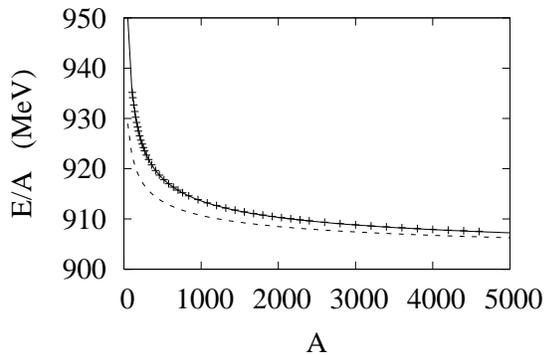}
\caption{Energy per baryon as a function of baryon number for
 $H\Lambda^2 = 2$ and $B^{1/4} = 156.26$ MeV. The exact results are
 indicated by the crosses, the fitted liquid-drop formula by the solid
 line. The dashed line corresponds to the liquid-drop formula without
 the curvature term.}
\label{fig:ea}
\end{figure}

As explained after Eq.~(\ref{betainfinite}), $(E/A)_{\mathit{bulk}}$
should be the energy per baryon of infinite matter with $\mu_e = 0$
rather than that of $\beta$ stable infinite matter. However, since we
consider only the CFL-type solution, this distinction is
irrelevant. Hence, for our chosen parameter sets, we have
$(E/A)_\mathit{bulk} = 900$ MeV. Since for the neutral strangelets the
Coulomb interaction has only a negligible effect on the total energies
(for example, in the case of the strangelets considered in
Section~\ref{sec:chdens}, the Coulomb interaction changes the total
energy per baryon by less than 5 keV) it will be neglected here in
order to reduce the numerical effort. The result of the fitted
coefficients $a_S$ and $a_C$ for the different parameter sets are
listed in Table~\ref{tab:liquiddrop}.
\begin{table}
\caption{\label{tab:liquiddrop} Fitted liquid-drop parameters for the
CFL-type neutral strangelets ($Z = 0$). The surface tension $\sigma$
corresponding to the fitted value of $a_S$ is also given.}
\begin{ruledtabular}
\begin{tabular}{ccccc}
$\begin{matrix}B^{1/4}\\\mbox{(MeV)}\end{matrix}$ &
$\begin{matrix}H\Lambda^2\\ \mbox{(MeV)}\end{matrix}$ &
$\begin{matrix}a_S\\ \mbox{(MeV)}\end{matrix}$ &
$\begin{matrix}a_C\\ \mbox{(MeV)}\end{matrix}$ &
$\begin{matrix}\sigma\\ \mbox{(MeV$/$fm$^2$)}\end{matrix}$
\\
\hline
156.26 & 2    & 107 & 289 & 11.9\\
159.46 & 2.25 & 109 & 297 & 12.8\\
163.46 & 2.5  & 112 & 306 & 13.9
\end{tabular}
\end{ruledtabular}
\end{table}
As an example, in order to show the accuracy of the asymptotic
expansion, we display in Fig.~\ref{fig:ea} some results for the energy
per baryon together with the liquid-drop formula,
Eq.~(\ref{liquiddrop}). 
The dashed line corresponds to the liquid-drop formula without the
curvature term ($a_C = 0$). From this figure it becomes clear that the
liquid-drop formula with curvature term works extremely well, much
better than in the case without pairing~\cite{GilsonJaffe}. The reason
is that shell effects are completely washed out because, contrary to
the situation in ordinary nuclei, the pairing gap is much larger than
the spacing between neighboring shells. Another interesting
observation is that the curvature term is very important, even for
rather large mass numbers $A$.

The coefficient $a_S$ is closely related to a very interesting
quantity, namely the surface tension. As explained in
Ref.~\cite{Parija}, the surface tension is obtained as
\begin{equation}
\sigma = \frac{E_S}{4\pi R_0^2}\,,
\label{defsigma}
\end{equation}
where
\begin{equation}
E_S = E - A\left(\frac{E}{A}\right)_\mathit{bulk}
\end{equation}
is the energy excess due to the surface and $R_0$ is an effective
radius defined by
\begin{equation}
A = \rho_{B\,\mathit{bulk}}\frac{4\pi R_0^3}{3}\,,
\end{equation}
which is actually very close to $R$ for not too small strangelets. On
the one hand, using the liquid-drop formula (\ref{liquiddrop}) for $E$
in Eq.~(\ref{defsigma}), one would obtain a surface tension which
depends on $A$ because of the curvature term. Therefore it is clear
that one has to use Eq.~(\ref{defsigma}) in the limit $A\to\infty$,
where the curvature term vanishes, i.e.,
\begin{equation}
\sigma = \frac{a_S\rho_{B\,\mathit{bulk}}^{2/3}}{(36\pi)^{1/3}}\,.
\end{equation}
The corresponding numbers are given in the last column of
Table~\ref{tab:liquiddrop}. They are of the same order of magnitude as
the estimate $\sigma \sim (70$ MeV$)^3 = 8.8$ MeV$/$fm$^2$ for SQM
without color superconductivity \cite{FarhiJaffe}. On the other hand,
the fact that the curvature term is very strong implies that the
knowledge of the surface tension alone might not be sufficient in
order to determine, e.g., the possibility of mixed phases, the size of
droplets, etc.

Before we conclude, let us comment on the physical meaning of the
surface tension we obtain. In the MIT bag model, it is supposed that
the energy needed to create a bag with volume $V$ is simply given by
$B V$. In principle one could imagine that there is an explicit
dependence of the bag energy on, e.g., the surface or the curvature of
the bag boundary. In Ref.~\cite{FarhiJaffe}, this contribution to the
surface tension was called ``intrinsic surface tension'', $\sigma_I$,
and it was argued that it should be small. What we calculate here is
the ``dynamical surface tension'', $\sigma_D$, which has its origin in
the change of the level density of the quarks inside the bag as a
function of the bag geometry.

\section{Summary}
\label{sec:sum}
In this paper we have investigated finite lumps of color
superconducting SQM. To that end we have treated the MIT bag model,
supplemented with a pairing interaction, in the framework of HFB
theory. This allows us to correctly include finite size effects for
pairing, too. The calculation is numerically rather involved, since in
addition to solving self-consistently the HFB equations, we have to
determine the bag radius and the fractions of the different quark
species by minimizing the total energy of the system.

As expected from previous MIT bag-model studies, we find a suppression
of the strange-quark densities at the surface, resulting in a positive
surface charge. Our main result is that, in spite of this surface
charge, the total charge of the CFL type solution is zero due to
pairing, as in bulk matter. Most of the positive surface charge is
compensated in a negatively charged layer situated at about 1-3 fm
below the surface. The origin of this concentration of the negative
charge is pairing: Since all quarks are paired, the positive surface
charge must be compensated on a length scale corresponding to the
coherence length. The remaining negative charge, which is necessary to
compensate all of the positive surface charge, is situated below this
layer. With increasing distance from the surface, the charge density
decreases on a length scale of $\sim 8$ fm, corresponding to the Debye
screening length. This number will probably be strongly decreased if
the gluons are included in a perturbative way similar to the photon.
In any case, in the biggest part of a large object, such as a strange
star, one finds vanishing charge density if one goes more than a few
tens of fm away from the surface. It remains to be investigated in
which way our results change the traditional picture of the surface of
a strange star and the detectability of smaller strangelets in current
experiments such as AMS-02 or LSSS~\cite{Madsen06}.

We have also compared our results for the energy per baryon of finite
strangelets with a liquid-drop like formula. We obtain a surface
tension of the order of 12-14 MeV, in reasonable agreement with
previous studies where color superconductivity was not considered, and
a strong curvature term which is crucial to reproduce the correct
energies up to baryon numbers of several thousands. An interesting
result is that, in the presence of color superconductivity, the
liquid-drop formula describes very accurately the total energies even
for $A\lesssim 100$, at least for strangelets with even baryon
number. The reason is that, since the gap $\Delta$ is much larger than
the spacing between the energy levels, shell effects are strongly
suppressed.
\section*{Acknowledgements}
We thank Michael Buballa for useful discussions and for the critical
reading of the first version of this manuscript.
\begin{appendix}
\section{Spinors in a spherical cavity}
\label{app:basis}
In this appendix we recall basic properties of free Dirac spinors in a
spherical cavity (see, e.g., Ref.~\cite{Bhaduri}). They can be written as
\begin{equation}
\psi_{fj\kappa mn}(\vek{r}) = \begin{pmatrix} 
g_{fj\kappa n}(r) \mathcal{Y}^m_{jl}(\Omega)\\ 
i\,f_{fj\kappa n}(r) \mathcal{Y}^m_{jl'}(\Omega)\end{pmatrix}\,,
\end{equation}
where $\mathcal{Y}$ are spinor spherical
harmonics~\cite{Varshalovich}. We have the following relations between
the angular momentum quantum numbers
\begin{alignat}{3}
&\kappa = j+\frac{1}{2}    &\quad\rightarrow\quad 
  l =  j+\frac{1}{2},\quad & l' = j - \frac{1}{2} \nonumber\\
&\kappa = -(j+\frac{1}{2}) &\quad\rightarrow\quad
  l =  j-\frac{1}{2},\quad & l' = j + \frac{1}{2}\,.
\end{alignat}
For the solutions of the free Dirac equation, the functions $f$ and
$g$ are given as follows in terms of the spherical Bessel functions
($\xi_{fj\kappa n} = \sqrt{p_{fj\kappa n}^2 + m_f^2}$)
\begin{align}
g_{fj\kappa n}(r) =& C_{fj\kappa n} \, j_l(p_{fj\kappa n}r) \nonumber\\
  f_{fj\kappa n} (r) =& C_{fj\kappa n} \mathrm{sgn}(\kappa n)
    \sqrt{\frac{\xi_{fj\kappa n} - m_f}{\xi_{fj\kappa n} + m_f}}\,
    j_{l'}(p_{fj\kappa n} r)\,,
\end{align}
where the $C_{fj\kappa n}$ are normalisation coefficients which can be
determined from the normalization
\begin{equation}
\int_0^R dr r^2 \int d\Omega \psi^\dagger(\vek{r}) \psi(\vek{r}) = 1\,.
\end{equation}
The momenta $p_{fj\kappa n}$ are obtained from the boundary condition. 
The boundary condition of the MIT bag model,
Eq.~(\ref{boundarycondition}), translates into the following equation
\begin{equation}
f_{fj\kappa n}(R) = -g_{fj\kappa n}(R)\,,
\end{equation}
or, explicitly,
\begin{equation}
j_l(p_{fj\kappa n} R) = \mathrm{sgn}(\kappa n) \sqrt{\frac{
    \xi_{fj\kappa n}- m_f}{\xi_{fj\kappa n} + m_f}
} \, j_{l'}(p_{fj\kappa n} R)\,,
\end{equation}
where we number by $n > 0$ the positive-energy (particle) states and
by $n < 0$ the negative-energy (antiparticle) states. In practice, we
will keep only the states with positive eigenvalues and neglect the
antiparticle contributions. The latter can approximately be absorbed
into a redefintion of the coupling constant.
\section{HFB equations}
\label{app:hfb}
In this appendix we will give some more details about the HFB
equations. Their derivation is analogous to the derivation of the
Dirac-Hartree-Bogoliubov equations in finite nuclei, which is given in
Ref.~\cite{CH}.

The HFB equations are derived from the Lagrangian by minimizing the
energy in the mean field approximation, i.e., linearizing the
interaction under the assumption of nonzero expectation values for the
condensates $s_{AA^\prime}(x)$, Eq.~(\ref{saa}). Due to the inhomogeneities
of a finite system, the Green's functions become nondiagonal in
momentum. In the stationary case, it is convenient to work in
$\vek{r}$ space for the spatial coordinates but to perform the Fourier
transformation for the time variable. Then the Green's functions,
\begin{equation} 
S(x,y) = -i \langle T(\Psi(x)\bar{\Psi}(y))\rangle~, 
\end{equation}
with
\begin{equation}
\Psi(x) = \begin{pmatrix} \psi(x)\\\psi_T(x)\end{pmatrix}
\end{equation}
take
the following general form in Nambu-Gorkov space:
\begin{multline}
S(\vek{r},\vek{r}^\prime;\omega) = \begin{pmatrix}
    G(\vek{r},\vek{r}^\prime;\omega)&  
    F(\vek{r},\vek{r}^\prime;\omega)\\
    \tilde{F}(\vek{r},\vek{r}^\prime;\omega)&  
    \tilde{G}(\vek{r},\vek{r}^\prime;\omega)\end{pmatrix}\\
  = 
\sum_{\alpha (\epsilon_\alpha >0)} \begin{pmatrix}
    U_\alpha(\vek{r})\\V_\alpha(\vek{r})\end{pmatrix}
    \frac{1}{\omega-\epsilon_\alpha + i\eta}
    (\bar{U}_\alpha(\vek{r}^\prime),\bar{V}_\alpha(\vek{r}^\prime))\\
  +   
\sum_{\beta (\epsilon_\beta < 0)} \begin{pmatrix}
    U_\beta(\vek{r})\\V_\beta(\vek{r})\end{pmatrix}
    \frac{1}{\omega+\epsilon_\beta - i\eta}
    (\bar{U}_\beta(\vek{r}^\prime),\bar{V}_\beta(\vek{r}^\prime))\,,
\label{greensfunctions}
\end{multline}
where $G,\tilde{G}$ and $F,\tilde{F}$ are normal and anomalous Green's
functions, respectively. The spinors $U_{\alpha,\beta}$ and
$V_{\alpha,\beta}$ correspond to the particle- and hole-like
components, respectively.

The energy in mean-field approximation can now be written as \cite{CH}
\begin{multline}
E_q = \int d^3 x \left(i\,
  \mathrm{Tr}[(i\vek{\gamma}\cdot\vek{\nabla}-m)
  G(x,x^+)]\phantom{\int}\right.
\\
  -\left.\frac{i}{2}\int d^4 y \,\mathrm{Tr}[\Sigma(x,y) G(y,x^+) - \Delta(x,y)
  \tilde{F}(y,x^+)]\right)~,
\label{energygeneral}
\end{multline}
where the derivative in the first term acts only on $x$ and not on
$x^+$, and $x^+$ means the four vector $(x^0+t,\vek{x})$ in the limit
$t \to 0^+$. In our case, the normal and anomalous self-energies
$\Sigma$ and $\Delta$ are local and time-independent:
$\Sigma(x,y) = eQA_0(\vek{x})\gamma^0\delta(x-y)$ and $\Delta(x,y) =
\Delta(\vek{x})\delta(x-y)$, and Eq.~(\ref{energygeneral}) can be
reduced to Eq.~(\ref{quarkenergy}).

As mentioned in Section~\ref{sec:hfb}, the expectation values (like
condensates, densities, etc.) which are needed for calculating
self-consistently the self-energy $\Sigma$ and the pairing field
$\Delta$ can be expressed in terms of the $U$ and $V$ functions. To
that end, it is sufficient to express them in terms of the Green's
functions, e.g.
\begin{equation}
s_{AA^\prime} = - \langle\bar{\psi}_T(x) \tau_A \lambda_{A^\prime}
  \psi(x)\rangle = i\, \mathrm{Tr}\, F(x,x^+) \tau_A \lambda_A\,,
\end{equation}
which leads to Eq.~(\ref{saadivergent}).

By minimizing the total energy with respect to the $U$ and $V$
functions, one obtains the HFB equations, [see
Eq.~(\ref{hfbequations})]:
\begin{equation}
\mathcal{H} W_\alpha = \epsilon_\alpha W_\alpha\,,
\end{equation}
with $W_\alpha = (U_\alpha,V_\alpha)^T$ and $\mathcal{H}$ being the
matrix on the left-hand side of Eq.~(\ref{hfbequations}).

For homogeneous infinite systems the matrix elements of $\mathcal{H}$
are diagonal in momentum space and solutions to the HFB equations are
known for many cases. For finite systems, in general, these equations
are solved numerically by diagonalizing the matrix $\mathcal{H}$ in
some conveniently chosen basis. Here, we are working in the basis
which diagonalizes the Dirac hamiltonian (i.e., $h_{fc}$ without the
Coulomb potential), see Appendix~\ref{app:basis}, and the eigenvectors
$U_\alpha(\vek{r})$ and $V_\alpha(\vek{r})$ are developped within this
basis.

The matrix elements of the pairing fields $\Delta_A(r)$ and of the
Coulomb field $A_0(r)$ are computed in the usual way. For
illustration, we give here the explicit expression for the matrix
elements of $\Delta_2(r)$, which connects up and down quarks, in the
basis described in Appendix~\ref{app:basis}:
\begin{multline}
(\Delta_2)_{j\kappa nn'} = \int_{r<R} d^3 r\,
  \psi^\dagger_{uj\kappa mn}(\vek{r})\Delta_2(r)\psi_{dj\kappa mn'}(\vek{r})\\
  = \int_0^R dr\, r^2 \Delta_2(r) (g_{uj\kappa n}(r) g_{dj\kappa n'}(r)\\
    +f_{uj\kappa n}(r) f_{dj\kappa n'}(r))\,.
\end{multline}
Note that, due to spherical symmetry, all matrices are diagonal in $j$
and $\kappa$ and proportional to the unit matrix with respect to
$m$. 

In spite of the spherical symmetry, the matrix to be diagonalized is
still huge, limiting the baryon number which can be calculated with
reasonable computational effort. It is therefore important to reduce
the size of the actual matrix to be diagonalized. By means of an
orthogonal transformation
\begin{equation}
\tilde{\mathcal{H}} = S \mathcal{H} S^T, \quad \tilde{W}
  = S W, \quad S S^T = 1
\end{equation}
in color, flavor, and Nambu-Gorkov space, the matrix can actually be
block-diagonalized (see, e.g., Ref.~\cite{ABR99}) containing seven
blocks. Six of them, $\tilde{\mathcal{H}}_{B,\dots G}$, are $2\times
2$ matrices in Nambu-Gorkov space, describing mutual pairing of two
particles, such as, e.g., red down quarks ($dr$) with green up quarks
($ug$):
\begin{equation}
\tilde{\mathcal{H}}_B = \begin{pmatrix} h_{ug} & \Delta_2 \\
  \Delta_2 & -h_{dr} \end{pmatrix}\,,
\end{equation}
where $h_{fc}$ is the single particle Hamiltonian for flavor $f$ and
color $c$. The second and third $2\times 2$ blocks are
\begin{equation}
\tilde{\mathcal{H}}_C = \begin{pmatrix} h_{ub} & \Delta_5 \\
  \Delta_5 & -h_{sr} \end{pmatrix}\,,\quad
\tilde{\mathcal{H}}_D = \begin{pmatrix} h_{db} & \Delta_7 \\
  \Delta_7 & -h_{sg} \end{pmatrix}\,.
\end{equation}
Since we have in addition the pairwise relations
$\tilde{\mathcal{H}}_{E,F,G} = -\tilde{\mathcal{H}}_{B,C,D}$, only
three of the six $2\times 2$ blocks have to be diagonalized in
practice. The seventh block, $\tilde{\mathcal{H}}_A$, is $6\times 6$
in Nambu-Gorkov space and describes pairing between red up, green down
and blue strange quarks
\begin{equation}
\tilde{\mathcal{H}}_A = \begin{pmatrix}
  h_{ur} &0&0&0&\Delta_2&\Delta_5\\
  0&h_{dg}&0&\Delta_2&0&\Delta_7\\
  0&0&h_{sb}&\Delta_5&\Delta_7&0\\
  0&\Delta_2&\Delta_5&-h_{ur}&0&0\\
  \Delta_2&0&\Delta_7&0&-h_{dg}&0\\
  \Delta_5&\Delta_7&0&0&0&-h_{sb}\end{pmatrix}\,.
\end{equation}

\section{Cutoff for the gap equation}
\label{app:cutoff}
As mentioned in Section~\ref{sec:hfb}, the divergent gap equation is
regularized with the help of a smooth cutoff function
\begin{equation}
f(p/\Lambda) = \frac{1}{1 + c_1 \exp(c_2 a (p/\Lambda - 1))}\,,
\end{equation}   
where $c_1 = \sqrt{2}-1$, $c_2 = 1/(4-2\sqrt{2})$, and $a = 22.58$
have been chosen such that $f^2(p/\Lambda)$ approximates the cutoff
function $g(p/\Lambda)$ used in Ref.~\cite{YasuiHosaka}, but our
function has the advantage to fall off more rapidly at very high
momenta, which allows us to truncate the basis at a lower energy.

This function is used as a form factor multiplying each of the four
legs of the four-point vertex. In practice, this means that the form
factor is used in two places: First, when calculating $s_{AA}(r)$, and
second, when calculating the matrix elements of $\Delta_A(r)$ in the
basis of the spinors defined in Appendix~\ref{app:basis}. It should be
noted that the diagonalization of the HFB matrix does not directly
provide us with the eigenfunctions $U_\alpha(\vek{r})$ and
$V_\alpha(\vek{r})$, but with their respective expansion coefficients
in the basis of the spinors defined in Appendix~\ref{app:basis}. When
calculating $s_{AA}(r)$ according to Eq.~(\ref{saadivergent}), the
coefficients have to be multiplied with the corresponding basis
functions, and in this step the factor $f(p_{fj\kappa n}/\Lambda)$ is
attached to each basis function. Second, when calculating the matrix
elements of the gap $\Delta_A$, we again attach a factor
$f(p_{fj\kappa n}/\Lambda)$ to each basis function.
\end{appendix}



\begin{thebibliography}{99}
\bibitem{colorsup} J.C. Collins and M.J. Perry, Phys. Rev. Lett. {\bf
  34}, 1353 (1975);
  B. Barrois, Nucl. Phys. \textbf{B129}, 390 (1977);
  S.C. Frautschi, Asymptotic freedom and color superconductivity in
  dense quark matter, in: Proc. of the Workshop on Hadronic Matter at
  Extreme Energy Density, N. Cabibbo (ed.), Erice 1978;
  D. Bailin and A. Love, Phys. Rep. \textbf{107}, 325 (1984).
\bibitem{Alford97} M.G. Alford, K. Rajagopal, and F. Wilczek,
  Phys. Lett. B \textbf{422}, 247 (1998).
\bibitem{Rapp97} R. Rapp, T. Sch\"afer, E.V. Shuryak, and
  M. Velkovsky, Phys. Rev. Lett. \textbf{81}, 53 (1998).
\bibitem{reviews} K. Rajagopal and F. Wilczek, in: M. Shifman
  (Editor), \textit{At the Frontier of Particle Physics, Handbook of
  QCD, Boris Ioffe Festschrift}, vol. 3, p. 2061 (World Scientific,
  Singapore 2001) [hep-ph/0011333];
  M. Alford, Ann. Rev. Nucl. Part. Sci. \textbf{51}, 131 (2001);
  T. Sch\"afer, hep-ph/0304281;
  D.H. Rischke, Prog. Part. Nucl. Phys. \textbf{52}, 197 (2004);
  M. Buballa, Phys. Rep. \textbf{407}, 205 (2005);
  H.-C. Ren, hep-ph/0404074;
  M. Huang, Int. J. Mod. Phys. E \textbf{14}, 675 (2005);
  I. A. Shovkovy, Found. Phys. \textbf{35}, 1309 (2005).
\bibitem{CFL} M.G. Alford, K. Rajagopal, and F. Wilczek,
  Nucl. Phys. \textbf{B 537}, 443 (1999).
\bibitem{weakCFL} I.A. Shovkovy and L.C.R. Wijewardhana, Phys. Lett. B
  \textbf{470}, 189 (1999);
  T. Sch\"afer, Nucl. Phys. \textbf{B 575}, 269 (2000);
  N.J. Evans, J. Hormuzdiar, S.D.H. Hsu, and M. Schwetz,
  Nucl. Phys. \textbf{B 581}, 391 (2000).
\bibitem{Witten} A.R. Bodmer, Phys. Rev. D \textbf{4}, 1601 (1971);
  E. Witten, Phys. Rev. D\textbf{30}, 272 (1984).
\bibitem{strangestars} C. Alcock, E. Farhi, and A. Olinto,
  Astrophys. J. \textbf{310}, 261 (1986);
  P. Haensel, J.L. Zdunik, and R. Schaeffer,
  Astron. Astrophys. \textbf{160}, 121 (1986).
\bibitem{cosmicrays} J. Madsen and J.M. Larsen,
  Phys. Rev. Lett. \textbf{90}, 121102 (2003);
  M. Rybczynski, Z. Wlodarczyk, and G. Wilk,
  Nucl. Phys. Proc. Suppl. \textbf{151}, 341 (2006).
\bibitem{Usov} V.V. Usov, Phys. Rev. Lett. \textbf{80}, 230 (1998).
\bibitem{UsovPage} D. Page and V.V. Usov, Phys. Rev. Lett.\textbf{89},
  131101 (2002).
\bibitem{nuggets} P. Jaikumar, S. Reddy and A.W. Steiner,
  Phys. Rev. Lett. \textbf{96},041101 (2006).
\bibitem{stablenuggets} M.G. Alford, K. Rajagopal, S. Reddy, and
  A.W. Steiner, Phys. Rev. D \textbf{73}, 114016 (2006).
\bibitem{RW01} K. Rajagopal and F. Wilczek,
  Phys. Rev. Lett. \textbf{86}, 3492 (2001).
\bibitem{Madsen01} J. Madsen, Phys. Rev. Lett. \textbf{87}, 172003
  (2001).
\bibitem{Usov04} V.V. Usov, Phys. Rev. D \textbf{70}, 067301 (2004).
\bibitem{MITbag} A. Chodos, R.L. Jaffe, K. Johnson, C.B. Thorn, and
  V.F. Weisskopf, Phys. Rev. D \textbf{9}, 3471 (1974).
\bibitem{Bhaduri} R.K. Bhaduri, \textit{Models of the Nucleon: From
  Quarks to Soliton} (Addison-Wesley, Redwood City 1988)
\bibitem{CH} B.V. Carlson and D. Hirata, Phys. Rev. C \textbf{62},
  054310 (2000).
\bibitem{FarhiJaffe} E. Farhi and R.L. Jaffe, Phys. Rev. D
  \textbf{30}, 2379 (1984).
\bibitem{LiuWilczek} W.V. Liu and F. Wilczek,
  Phys. Rev. Lett. \textbf{90}, 047002 (2003).
\bibitem{Abuki} H. Abuki, T. Hatsuda, K. Itakura, Phys. Rev. D
  \textbf{65}, 074014 (2002).
\bibitem{Tatsumi} T. Tatsumi, M. Yasuhira, and D.N. Voskresensky,
  Nucl. Phys. \textbf{A718}, 359c (2003).
\bibitem{GarciaMoliner} F. Garc\'{\i}a-Moliner and F. Flores,
  \textit{Introduction to the theory of solid surfaces}
  (Cambridge University Press, Cambridge, 1979).
\bibitem{SWR04} A. Schmitt, Q. Wang and D.H. Rischke, Phys. Rev. D
  \textbf{69}, 094017 (2004).
\bibitem{ABR00} M.G. Alford, J. Berges, and K. Rajagopal, Nucl. Phys.
  \textbf{B 571}, 269 (2000)
\bibitem{LitimManuel} D.F. Litim and C. Manuel, Phys. Rev. D \textbf{64},
  094013 (2001).
\bibitem{GilsonJaffe} E.P. Gilson and R.L. Jaffe,
  Phys. Rev. Lett. \textbf{71}, 332 (1993).
\bibitem{Parija} B.C. Parija, Phys. Rev. C \textbf{48}, 2483 (1993).
\bibitem{Madsen06} J. Madsen, arXiv:astro-ph/0612784 (2006).
\bibitem{Varshalovich} D.A. Varshalovich, A.N. Moskalev, and
  V.K. Khersonskii, \textit{Quantum Theory of Angular Momentum} (World
  Scientific, Singapore 1988).
\bibitem{ABR99} M.G. Alford, J. Berges, and K. Rajagopal,
  Nucl. Phys. B \textbf{558}, 219 (1999).
\bibitem{YasuiHosaka} S. Yasui and A. Hosaka, Phys. Rev. D
  \textbf{74}, 054036 (2006).
\end{thebibliography}
\end{document}